\documentclass[sigconf,natbib=false]{acmart}

\usepackage{tabularx}

\setcopyright{acmlicensed}
\copyrightyear{2026}
\acmYear{2026}
\acmDOI{XXXXXXX.XXXXXXX}
\acmISBN{978-1-4503-XXXX-X/2018/06}

\acmConference[AI Summit '26]{ACM AI Leadership Summit 2026}{June 2026}{Atlanta, GA}

\RequirePackage[
  datamodel=acmdatamodel,
  style=acmnumeric,
  sorting=none,
]{biblatex}

\begin{document}

\title{Toward Exascale AI for Science: A Scalable AI Skill for Autonomous Microkinetics Discovery}

\author{Ken-ichi Nomura}
\email{knomura@usc.edu}
\orcid{0000-0002-1743-1419}
\affiliation{%
  \institution{University of Southern California}
  \city{Los Angeles}
  \state{CA}
  \country{USA}
}

\author{William Dawson}
\email{william.dawson@riken.jp}
\orcid{0000-0003-4480-8565}
\affiliation{%
  \institution{RIKEN R-CCS}
  \city{Kobe}
  \country{Japan}
}

\author{Kai Ito}
\email{ki\_790@usc.edu}
\orcid{0009-0006-7260-7807}
\affiliation{%
  \institution{Kumamoto University}
  \city{Kumamoto}
  \country{Japan}
}

\author{Nabankur Dasgupta}
\email{ndasgupt@usc.edu}
\orcid{0000-0002-5262-6086}
\affiliation{%
  \institution{University of Southern California}
  \city{Los Angeles}
  \state{CA}
  \country{USA}
}

\author{Taufeq Mohammed Razakh}
\email{razakh@usc.edu}
\orcid{0000-0002-4540-0943}
\affiliation{%
  \institution{University of Southern California}
  \city{Los Angeles}
  \state{CA}
  \country{USA}
}

\author{Thomas Linker}
\email{tlinker@slac.stanford.edu}
\orcid{0000-0002-0504-4876}
\affiliation{%
  \institution{SLAC National Accelerator Laboratory}
  \city{Menlo Park}
  \state{CA}
  \country{USA}
}

\author{Aiichiro Nakano}
\email{anakano@usc.edu}
\orcid{0000-0003-3228-3896}
\affiliation{%
  \institution{University of Southern California}
  \city{Los Angeles}
  \state{CA}
  \country{USA}
}

\renewcommand{\shortauthors}{Nomura et al.}

\begin{abstract}
We present a scalable AI-driven framework that advances autonomous scientific discovery by combining agentic workflow automation, high-performance computing, and scientific surrogate models. Using microkinetics discovery as a testbed, the work demonstrates how AI can reduce expert intervention, recover from failed simulations, and systematically evaluate surrogate model reliability. This study shows how AI skills can transform complex domain workflows into robust, scalable capabilities for next-generation materials research.
\end{abstract}



\maketitle


\section{Introduction}

\begin{figure*}[t]
  \centering
  \includegraphics[width=\textwidth]{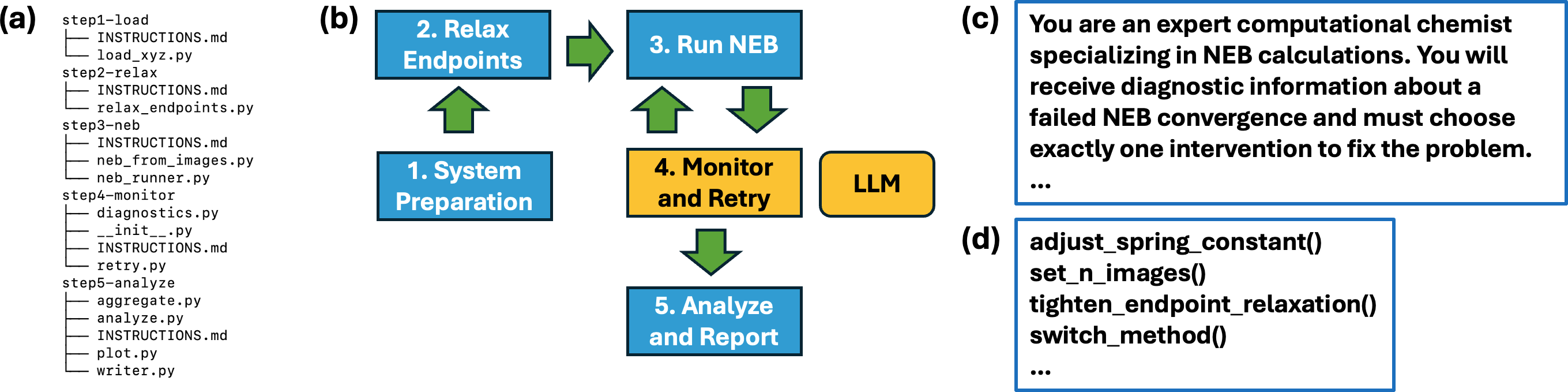}
  \caption{Design of NEB skill. (a) Directory and file organization of the five key steps. (b) Skill flowchart. A user initiates the skill by providing reactant and product coordinates. Locally executed tasks are indicated by blue boxes, while yellow boxes indicate HPC inference by LLM. During the \textit{Monitor and Retry} step, an agent is given the system task (c), and decides strategies (d) to improve the success rate of NEB calculation. }
  \label{fig:nebskill}
  \Description{Schematic of NEB skills.}
\end{figure*}

A grand challenge in science is to understand complex materials from first principles -- a challenge best addressed by multiscale simulation \cite{nobelprize01}. Today, the state of the art (SOTA) is combined deployment of multi-exaflop quantum mechanical (QM) simulations with surrogate foundation models \cite{razakh2025multiscale, hattori2025beyond}, a formidable modeling task at the intersection of high-performance computing (HPC), AI, and applied science. A prototypical use case is reaction kinetics of complex material structures \cite{doi:10.1126/science.aeb9934, doi:10.1021/acs.iecr.9b04625}, a regime where even small scale QM simulations are laborious and the establishment of robust foundation models remains unrealized. 





Agentic AI -- characterized by autonomy, planning, and task execution -- promises to vault materials research over these technical barriers through the automation of workflows that once only humans could manage. Agents deployed for scientific tasks integrate domain-specific knowledge and tools for long-term memory or information retrieval~\cite{manning2008vector, han2023vectorDBsurvey}, as well as interaction with external software~\cite{wang2025dreams, gao2025democratizing} or resources~\cite{Boiko2023Autonomous, Bran2024}. Composable augmentation is typically achieved today through two means: the Model Context Protocol (MCP)~\cite{anthropic_mcp_2024} and agent skills~\cite{agent_skills_spec_2026, vercel_skills_sh_2026}. Skills in particular have gained traction as a lightweight abstraction layer to teach Large Language Models (LLMs) domain-specific and procedural knowledge. 
Embedding these capabilities into agent harnesses transforms general-purpose frameworks (e.g., OpenClaw~\cite{steinberger_openclaw_2026}, NemoClaw~\cite{nvidia_nemoclaw_2026}) into "AI scientists." With the rapid adoption of automated AI workflows across the scientific community \cite{PyzerKnapp2022Accelerating, Boiko2023Autonomous}, ensuring their consistency, reproducibility, and reliability has become an urgent priority \cite{Lu2026Automation}.

Agent tools and skills to leverage HPC resources would accelerate materials research through three complementary pillars: agentic automation, materials-specific surrogate models, and large-scale simulations on exascale computers. Agents backed by open models running on national supercomputing infrastructure would in turn manage simulations on the same or complementary machines. Unfortunately, domain-specific frameworks -- particularly incorporating HPC~\cite{dawson2026lara} -- able to robustly perform simulations despite the sycophantic nature of LLMs~\cite{MacKnight2025}, remain in their infancy. In this work, we realize this vision as an HPC-centric agent able to perform reaction kinetics simulations at scale, and evaluate its performance for comparing different universal models for material properties. 




\section{Methodology}
\subsection{Skill Design}

To explore the potential of agent skills, we have developed ``nebskill''~\cite{nomuraNebskill} for Nudged Elastic Band (NEB)~\cite{jonsson1998neb, henkelman2000climbing_neb} calculations. NEB searches for the minimum energy pathway (MEP) between a reactant and product to determine the transition state and activation energy of a chemical reaction. Although the methodology is well established, NEB calculations are highly sensitive to the choice of parameters and the system of interest. This dependency often turns conceptually simple work into labor-intensive jobs that take days, if not weeks. Hence, agentic automation offers the promise of increasing scientific productivity for these studies.
Though fully autonomous AI scientists have garnered significant attention~\cite{Lu2026Automation}, their widespread adoption remains hindered by the lack of one-size-fits-all solutions; we therefore adopt a co-scientist approach for the ''nebskill'', emphasizing immediate, practical value for a community increasingly exploring agentic workflows. 

Figure \ref{fig:nebskill} (a) shows the organization of directories and files. The top directory contains a SKILL.md markdown file and each subdirectory contains another markdown file INSTRUCTIONS.md. This allows the agent to process the skill as it progresses, as well as to divide the entire task into smaller subtasks to reduce the chance of context rot \cite{hong2025context}. 
The five key steps of the skill are shown below:

\begin{itemize}
\item  \textbf{Step1}: Initialization: setup environment (e.g., download MLIP models, verify reactant and product coordinates). 

\item \textbf{Step2}: Relax reactant and product configurations.

\item \textbf{Step3}: Interpolate between the reactant and product, then run a two-phase calculation -- standard NEB followed by Climbing-Image (CI) NEB. 

\item \textbf{Step4}: Monitoring and failure recovery. Upon a calculation failure, LLM diagnoses the cause and proposes a new set of parameters (e.g., number of NEB images, spring constant). 

\item \textbf{Step5}: Analysis and report. Methods from domain experts are applied to converged calculations to verify the chemistry. 

\end{itemize}

When a job fails, the agent diagnoses the cause and proposes a new job to try (Step 4) following the ReAct (reasoning and acting) \cite{yao2023react} approach.
Based on observations from the failed job, the agent makes a decision on whether to exploit accumulated knowledge of the ongoing campaign (e.g., slightly update the value of spring constant) or to explore a new strategy (e.g., change the interpolation scheme from linear to spline). 
In this work, we use the ALCF (Argonne Leadership Computing Facility) Inference Endpoint, through which we access high-throughput AI services at scale. The API is called with a system prompt (Fig.~\ref{fig:nebskill}c) along with the current NEB parameters and the diagnosis to propose a next strategy (Fig.~\ref{fig:nebskill}d). 
Building on our previous integration of agents with supercomputer Fugaku~\cite{dawson2026lara}, we performed scalability tests on the Perlmutter cluster at the National Energy Research Scientific Computing Center (NERSC). Under the US-Japan Genesis partnership, we plan to use these globally distributed supercomputers as a single metacomputer to perform grand challenge AI-simulation workflows otherwise impossible, leveraging our previous US-Japan metacomputing paradigm~\cite{4090197}.

\subsection{Surrogate Models and Dataset}

Equivariant networks for interatomic potentials encode atomic environments by expanding interatomic displacements into spherical harmonics, yielding internal features that transform as proper geometric tensors under rotation \cite{Drautz2019atomic, batatia2025design}. 
By enforcing equivariance throughout network architecture, the model does not need to relearn the same physics in arbitrarily rotated orientations, yielding substantially better data efficiency, smoother and more physically consistent force/stress predictions, and improved extrapolation to geometries and compositions outside the training distribution. 

While many network architectures have been proposed to date \cite{riebesell2023matbench}, NequIP \cite{Batzner2022NequIP} and MACE \cite{Batatia2022MACE} are widely used equivariant MLIP models based on message-passing scheme to build multi-body correlations over successive graph convolutions. The Allegro model \cite{Musaelian2023Allegro} employs a strictly local feature approach where pairwise equivariant feature tensors are iteratively refined through tensor products with neighboring pairs sharing a central atom, preserving algorithmic scalability without sacrificing expressivity. Beyond these models supported by the AI skill, we expanded the evaluation pool to include a total of 12 MLIPs.

\begin{figure}[ht]
  \centering
  \includegraphics[width=\linewidth]{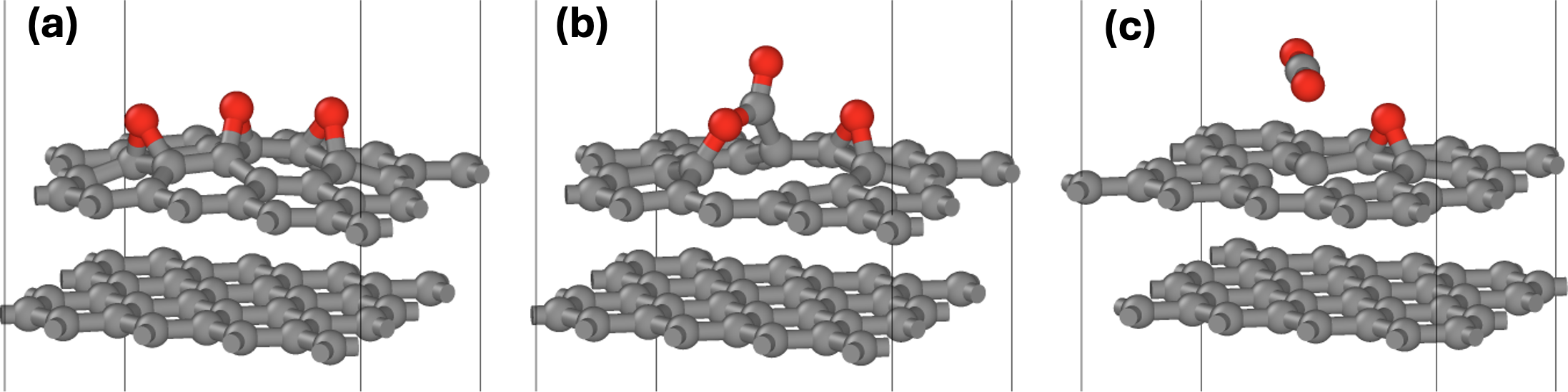}
  \caption{(a) Atomic configurations of reactant, (b) transition state, and (c) product obtained with ab-initio calculation. Red spheres for oxygen and gray ones for carbon atoms, respectively. The first step of NEB calculation is to smoothly interpolate the reactant and product configurations and generate an initial guess of intermediate states (i.e. NEB images).}
  \label{fig:nebconfigs}
  \Description{NEB results using DFT and various SOTA models.}
\end{figure}

In this work, the agent examines a number of SOTA pretrained universal MLIP models as a surrogate of graphite surface microkinetics. We evaluate over ten SOTA MLIP models to find the model that best matches with a given ground truth result.  
Figure~\ref{fig:nebconfigs} presents the atomic configurations during the sublimation of CO$_2$ from a graphite surface obtained by ab-initio quantum mechanical calculations and used to evaluate the developed AI skill. 

\textbf{QM calculation:} We perform adiabatic quantum molecular dynamics simulations using the QXMD software~\cite{ref:SHIMOJO2019100307}. A QMD simulation~\cite{ref:CarPhysRevLett.55.2471, ref:PayneRevModPhys.64.1045} follows the trajectory of all atoms, while calculating interatomic forces quantum mechanically from first principles within the framework of density functional theory (DFT)~\cite{ref:HohenbergPhysRev.136.B864}. The electronic states are calculated using the projector-augmented wave (PAW) method~\cite{ref:BlochlPhysRevB.50.17953}, where projector functions are generated for 2$s^{2}$2$p^{2}$ for C, $2s^{2}2p^{4}$ for O atoms. The PBE generalized gradient approximation (GGA) is used for the exchange-correlation functional~\cite{ref:PerdewPhysRevLett.77.3865} and DFT-D2 for the dispersion correction ~\cite{ref:GrimmeJCC2512}. The plane-wave cutoff energies are set to 30 Ry for the wavefunction and 300 Ry for the electron density in all calculations. We used $4 \times 4 \times 1$ $k$-points for Brillouin zone sampling.

\textbf{Graphite model:} The graphite model consists of two graphene layers. The graphene sheet consists of $4 \time 2$ unit cells with a lattice constant of 1.43 \AA. To prevent interaction between periodic images along the $z$ direction due to the periodic boundary condition, a vacuum layer of thickness 14 \AA\ is introduced along the $z$ axis. The supercell is orthorhombic with dimensions of $8.38 \times 9.91 \times 17.7$ \AA$^3$.

\section{Results}

\subsection{Reaction Coordinate and Activation Energy}

\begin{figure}[h]
  \centering
  \includegraphics[width=0.75\linewidth]{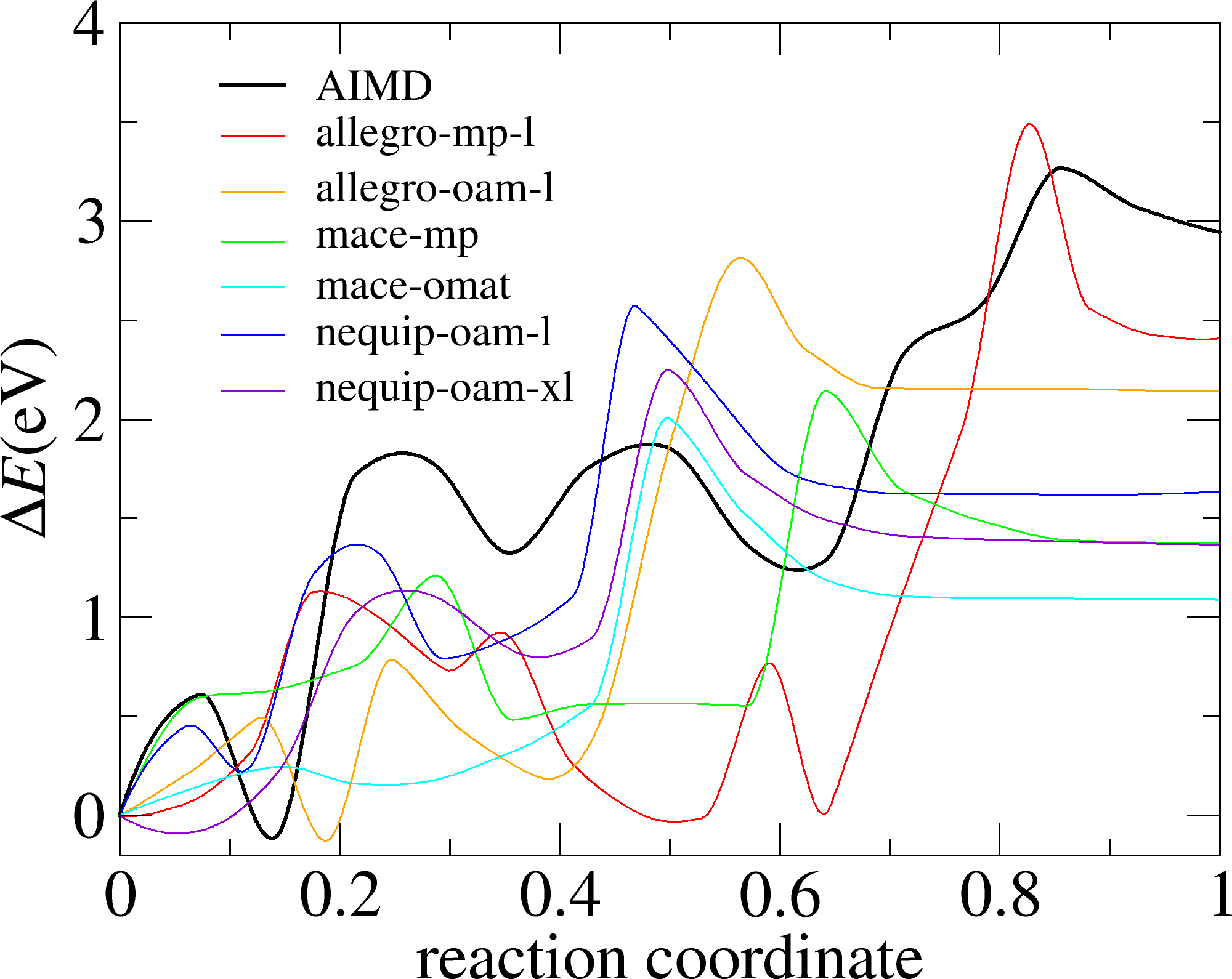}
  \caption{Activation energy along the reaction coordinate predicted by the default pretrained MLIP models compared against ground truth QM calculations (black). The horizontal axis is normalized by the total number of NEB images.}
  \label{fig:nebresults}
  \Description{Scalability test.}
\end{figure}

Fig. \ref{fig:nebresults} presents the energy profiles obtained by SOTA MLIP models together with DFT calculation as the ground truth. We evaluated pretrained universal MLIPs spanning the MACE, NequIP, Allegro, DPA-2/DPA-3, EquiformerV3, PET-MAD and MatRIS models \cite{Batatia2025MACEFoundation, Batzner2022NequIP, Musaelian2023Allegro, zhang2024dpa2, zhang2025dpa3, liao2026equiformerv3, mazitov2025petmad, zhou2026matris}. While quantitative comparisons may not be straightforward without fine-tuning, some of the MLIP models agree reasonably well with the ground truth calculation. For example, among all models, mace-mp-l shows the closest activation energy, while the energy curve of nequip-oam and allegro-oam closely follow the DFT result. In terms of the reaction mechanism, all models predict the lactone-mediated CO$_2$ sublimation mechanism that was also verified by other QM calculations~\cite{ref:SunJPC115,ref:LarcipreteJACS133}.

\subsection{Scalability Test}

To maximize parallel efficiency and improve time-to-solution, our agent skill employs two-dimensional parallelism over the MLIP model and NEB images. While the concurrency of model parallelism differs case by case, typically 5 to 10 MLIP models are evaluated simultaneously to improve the statistics. Since the model parallelism is loosely coupled, each MLIP evaluation is distributed over different nodes. In contrast, parallelization over NEB images is tightly coupled due to frequent atomic force exchanges. To minimize communication latency, the NEB images (typically 10 to 30) are distributed across the GPUs of a single node in a round-robin fashion to accelerate the overall calculation.

We have performed scalability tests of the AI skill on the Department of Energy (DOE) leadership computing facilities -- specifically the Perlmutter cluster at NERSC and the Inference Endpoint at ALCF. The Perlmutter cluster is an HPE Cray system consisting of 3,072 CPU and 1,792 GPU nodes. Each GPU node hosts four NVIDIA A100 GPUs. The ALCF Inference Endpoint is powered by the Sophia (NVIDIA DGX A100) and Metis (SambaNova) clusters, and offers DOE scientists high-throughput, low-latency AI inference capabilities. Currently dozens of open foundation models are supported, including Google Gemma \cite{gemmateam2024gemma}, Meta LLaMA \cite{dubey2024llama3}, and OpenAI GPT-OSS \cite{openai2025gptoss}. 

Table \ref{table:scaling} shows the improvement in runtime from baseline. All measurements were performed on Perlmutter at NERSC. While the scalability performance is impacted by many external factors such as file system access and network congestion, we have achieved a nearly perfect parallel efficiency, i.e. ~3.75$\times$ speedup on four GPU nodes.

\begin{table}[h]
\caption{Scalability of distributed NEB computing }
\label{table:scaling}
\begin{tabular}{c c c c}
 \hline
Node & GPUs & wallclock (min) & speed up \\
 \hline \hline
1 & 4 & 116.9 & 1.0 (baseline) \\
2 & 8 & 84.3 & 1.39 \\
3 & 12 & 61.0 & 1.92 \\
4 & 16 & 32.1 & 3.75 \\
 \hline
\end{tabular}
\end{table}

\subsection{Failure and Retry Analysis}
Because NEB calculations explore the energy landscape of target system, its success is highly dependent on researchers' knowledge and familiarity of the target system. Researchers often figure out an optimal configuration after time-consuming trial-and-error efforts. Therefore, investigating the cause of NEB calculation failures, retry strategies, and its cost would provide useful insights of the system and actionable strategies for improvement. Here, we present a case study on failed NEB jobs and the robustness of MLIP model predictions. 

We have identified 54 jobs in this study that failed in the first attempt but successfully converged at the end. We call them "resilient" NEB job. Of all, 51 jobs exhibited “image collapse” where the distance between each NEB image became too close, most likely attributed to the value of spring constant being too low. The remaining three jobs are “kinking” in the energy profile, exhibiting discontinuities or being very noisy.

\begin{table}[h]
\caption{Five representative NEB calculations that successfully converged after several retries by the agent. Used MLIP model, attempted retries (from top to bottom), the number of tokens consumed, and communication time with ALCF Inference Endpoint are shown respectively.  }
\label{table:nebsuccess}
\begin{tabularx}{\linewidth}{X X X l}
\hline
MLIP & retry & tokens & comm. (s) \\
\hline\hline
mace-mp & 
adjust spring \newline 
adjust spring \newline
set N images \newline 
adjust spring & 6642 & 23.17 \\
\hline
allegro-oam-l & adjust spring & 1595 & 14.48 \\
\hline 
pet-omat & adjust spring \newline
adjust spring \newline 
set N images & 4686 & 19.43 \\
\hline
pet-omat & adjust spring \newline
adjust spring & 3305 & 11.03\\
\hline
allegro-oam-l & adjust spring & 1593  & 14.41 \\
\hline

\end{tabularx}
\end{table}

Table \ref{table:nebsuccess} presents five representative cases out of the resilient job pool. The default strategy taken by the agent turned out to be adjusting the spring constant. While many jobs were successfully converged by one or a few attempts of tightening the spring constant, several resilient jobs went through multi-step parameter adjustments. An example is mace-mp shown in Table \ref{table:nebsuccess} where the agent chose to increase the spring constant twice consecutively, i.e. $0.3 \rightarrow 0.5 \rightarrow$ 1.2, then increased the NEB image size by 2 from 13. Finally, the agent decreased the spring constant from 1.2 $\rightarrow$ 0.5 and successfully achieved the energy convergence. Token usage averaged approximately 1,600 per retry, with communication times ranging from 6 to 14 seconds, signifying sound throughput performance across the DOE leadership computing facilities.

AI-driven workflow automation can also validate the robustness of scientific discoveries. Table \ref{table:table_ea} shows the sensitivity of MLIP prediction by adjusting a convergence criterion; the maximum force between 0.1 - 0.3 eV/\r{A}. While the mean is an important indicator to select a suitable surrogate from a dozen of very powerful pretrained MLIP models currently available, highly accurate model prediction does not guarantee its robustness \cite{Fuetal2022}. The distribution of the model predictions helps researchers to systematically select a set of MLIP models suitable for microkinetics studies.

\begin{table}[t]
\caption{Sensitivity of activation energy with NEB parameters}
\label{table:table_ea}
\begin{tabular}{c c c c c}
 \hline
MLIP&mean&min&max&stdev \\
\hline \hline
allegro-mp-l&3.3425&3.3064&3.5144&0.0615 \\
allegro-oam-l&3.0453&2.812&3.1303&0.1226 \\
dpa2&2.2842&2.1733&2.5763&0.1145 \\
dpa3&2.5555&2.5128&2.5982&0.0438 \\
equiformer-v3&2.153&2.1523&2.1543&0.0008 \\
mace-mp&2.0936&1.9454&2.146&0.0529 \\
mace-omat&1.7949&1.7676&2.0726&0.0667 \\
matris-oam&2.9847&2.8798&3.3752&0.1337 \\
nequip-mp-l&1.476&1.3604&2.0854&0.1602 \\
nequip-oam-l&2.2994&2.2158&2.6137&0.0895 \\
nequip-oam-m&2.4265&2.4244&2.431&0.0018 \\
pet-omat&2.0713&2.047&2.2021&0.0535 \\
  \hline
\end{tabular}
\end{table}

\section{Conclusion}
We have developed a portable and scalable AI skill that automates NEB calculations by combining agentic reasoning, HPC, and universal MLIPs. Beyond reducing manual intervention, our study demonstrates how domain expertise can be encapsulated into reusable AI skills that autonomously recover from failures, benchmark competing surrogate models, and quantify the robustness of scientific predictions. AI automation also facilitates uncertainty quantification; a critical step for reliable and reproducible science.
As scientific foundation models and HPC resources continue to grow, collections of interoperable AI skills executed across distributed computing facilities could form the basis of autonomous scientific campaigns capable of accelerating materials discovery in the era of exascale computing and beyond.


\begin{acks}
The authors acknowledge support from the ARO cooperative agreement W911NF-25-2-0183. This research used computing resources under Innovative and Novel Computational Impact on Theory and
Experiment (INCITE) at the Argonne Leadership Computing Facility. This research also used resources of the National Energy Research Scientific Computing Center (NERSC), a Department of Energy User Facility using NERSC award BES-ERCAP 0035639. This work used computational resources of Hokusai provided by RIKEN through the HPCI System Research Project (Project ID: hp260089).
\end{acks}

\printbibliography

@software{steinberger_openclaw_2026,
  author = {Peter Steinberger and {The OpenClaw Foundation}},
  title = {OpenClaw: An Open-Source Autonomous AI Agent Framework},
  year = {2026},
  url = {https://github.com/OpenClaw/openclaw},
  note = {Autonomous personal AI agent framework; accessed June 18, 2026}
}

@software{nvidia_nemoclaw_2026,
  author = {{NVIDIA Corporation}},
  title = {NVIDIA NemoClaw: Enterprise-Grade Agent Orchestration and Security Stack},
  year = {2026},
  url = {https://blogs.nvidia.com/blog/industrial-software-leaders-secure-autonomous-ai-engineers-nemoclaw/},
  note = {Enterprise secure distribution wrapper for OpenClaw; accessed June 18, 2026}
}

@online{anthropic_mcp_2024,
  author = {{Anthropic} and {Agentic AI Foundation}},
  title = {Model Context Protocol (MCP): An Open Standard for Integrating AI Models with External Tools and Data},
  year = {2024},
  url = {https://modelcontextprotocol.io},
  urldate = {2026-06-18},
  %note = {Maintained by the Agentic AI Foundation under the Linux Foundation}
}

@book{manning2008vector,
  author    = {Manning, Christopher D. and Raghavan, Prabhakar and Sch\"{u}tze, Hinrich},
  title     = {Introduction to Information Retrieval},
  publisher = {Cambridge University Press},
  year      = {2008},
 isbn = {0521865719},
 address = {New York}
}

@article{han2023vectorDBsurvey,
  author    = {Han, Yikang and Liu, Ai and Wang, Haihao},
  title     = {A Survey of Vector Databases in the Era of Large Language Models},
  journal   = {arXiv preprint arXiv:2310.11703},
  year      = {2023},
  %note      = {Standard survey for modern vector database architectures}
}

@incollection{jonsson1998neb,
  author    = {J\'{o}nsson, Hannes and Mills, Greg and Jacobsen, Karsten W.},
  title     = {Nudged Elastic Band Method for Finding Minimum Energy Paths of Transitions},
  booktitle = {Classical and Quantum Dynamics in Condensed Phase Simulations},
  editor    = {Berne, Bruce J. and Ciccotti, Giovanni and Coker, David F.},
  publisher = {World Scientific},
  pages     = {385--404},
  year      = {1998},
  %note      = {The original seminal paper formulating the standard NEB algorithm}
}

@article{henkelman2000climbing_neb,
  author    = {Henkelman, Graeme and Uberuaga, Blas P. and J\'{o}nsson, Hannes},
  title     = {A Climbing Image Nudged Elastic Band Method for Finding Saddle Points and Minimum Energy Paths},
  journal   = {The Journal of Chemical Physics},
  volume    = {113},
  number    = {22},
  pages     = {9901--9904},
  year      = {2000},
  doi       = {10.1063/1.1329672}
}

@online{agent_skills_spec_2026,
  author    = {{Agent Skills Working Group}},
  title     = {Agent Skills Specification: Portable AI Agent Expertise via SKILL.md},
  year      = {2026},
  url       = {https://agentskills.io},
  urldate   = {2026-06-18},
 % note      = {The official standard for progressive-disclosure procedural capabilities across AI coding agents}
}

@software{vercel_skills_sh_2026,
  author    = {{Vercel Labs} and {The Agent Skills Directory Contributors}},
  title     = {The Agent Skills Directory and CLI (skills.sh)},
  year      = {2026},
  url       = {https://www.skills.sh},
  %note      = {The central registry and open ecosystem package manager (`npx skills`) for installing reusable agent capabilities}
}

@techreport{hong2025context,
  author      = {Hong, Kelly and Troynikov, Anton and Huber, Jeff},
  title       = {Context Rot: How Increasing Input Tokens Impacts LLM Performance},
  institution = {Chroma},
  date        = {2025-07},
  url         = {https://research.trychroma.com/context-rot},
  %note        = {Comprehensive empirical study mapping performance cliffs across 18 frontier long-context language models}
}

@software{nomuraNebskill,
  author       = {Nomura, Ken-ichi},
  title        = {nebskill: Nudged Elastic Band software utilities and analysis tools},
  year         = {2026},
  publisher    = {GitHub},
  journal      = {GitHub repository},
  url = {https://github.com/KenichiNomura/nebskill},
  commit       = {main},
  lastaccessed = {Jun 22, 2026}
}

@article{gemmateam2024gemma,
  author    = {{Gemma Team} and Mesnard, Thomas and Hardin, Rustin and Dada, Shreya and Bhupatiraju, Surya and Laurent, S{\'e}bastien and Almasi, Th{\'e}o and Ye, Zouye and Cao, Yuan and Hurt, Nathan and others},
  title     = {Gemma: Open Models Based on Gemini Research},
  year      = {2024},
  eprint    = {2403.08295},
  archivePrefix = {arXiv},
  url       = {https://arxiv.org/abs/2403.08295}
}

@article{dubey2024llama3,
  author    = {Dubey, Abhimanyu and Jauhri, Abhinav and Pandey, Abhinav and Kadian, Agam and Al-Dahle, Ahmad and Letman, Aiesha and Akhoundi, Akhil and Alisankus, Alon and Anubhai, Rishabh and Balaji, Prajjwal and others},
  title     = {The LLaMA 3 Herd of Models},
  year      = {2024},
  eprint    = {2407.21783},
  archivePrefix = {arXiv},
  primaryClass  = {cs.AI},
  url       = {https://arxiv.org/abs/2407.21783}
}

@misc{openai2025gptoss,
  author       = {{OpenAI}},
  title        = {gpt-oss: Open-weight language models with advanced reasoning and tool use capabilities},
  year         = {2025},
  publisher    = {GitHub},
  journal      = {GitHub repository},
  howpublished = {\url{https://github.com/openai/gpt-oss}},
  commit       = {main}
}

@inproceedings{yao2023react,
  author    = {Yao, Shunyu and Zhao, Jeffrey and Yu, Dian and Du, Nan and Shafran, Izhak and Narasimhan, Karthik R. and Cao, Yuan},
  title     = {{ReAct}: Synergizing Reasoning and Acting in Language Models},
  booktitle = {International Conference on Learning Representations (ICLR)},
  year      = {2023},
  publisher = {OpenReview.net},
  url       = {https://openreview.net/forum?id=WE_vluYUL-X}
}

@article{Bran2024,
author={M. Bran, Andres
and Cox, Sam
and Schilter, Oliver
and Baldassari, Carlo
and White, Andrew D.
and Schwaller, Philippe},
title={Augmenting large language models with chemistry tools},
journal={Nat. Mach. Intell.},
date={2024-05},
volume={6},
number={5},
pages={525-535},
issn={2522-5839},
doi={10.1038/s42256-024-00832-8},
url={https://doi.org/10.1038/s42256-024-00832-8}
}

@article{gao2025democratizing,
  title={Democratizing AI scientists using ToolUniverse},
  author={Gao, Shanghua and Zhu, Richard and Sui, Pengwei and Kong, Zhenglun and Aldogom, Sufian and Huang, Yepeng and Noori, Ayush and Shamji, Reza and Parvataneni, Krishna and Tsiligkaridis, Theodoros and others},
  journal={arXiv:2509.23426},
  year={2025}
}

@article{wang2025dreams,
  title={DREAMS: Density functional theory based research engine for agentic materials simulation},
  author={Wang, Ziqi and Huang, Hongshuo and Zhao, Hancheng and Xu, Changwen and Zhu, Shang and Janssen, Jan and Viswanathan, Venkatasubramanian},
  journal={arXiv:2507.14267},
  year={2025}
}

@Article{Lu2026Automation,
author={Lu, Chris
and Lu, Cong
and Lange, Robert Tjarko
and Yamada, Yutaro
and Hu, Shengran
and Foerster, Jakob
and Ha, David
and Clune, Jeff},
title={Towards end-to-end automation of AI research},
journal={Nature},
date={2026-03},
volume={651},
number={8107},
pages={914-919},
issn={1476-4687},
doi={10.1038/s41586-026-10265-5},
url={https://doi.org/10.1038/s41586-026-10265-5}
}

@article{dawson2026lara,
  title={LARA: Validation-Driven Agentic Supercomputer Workflows for Atomistic Modeling},
  author={Dawson, William and Beal, Louis and Cur{\'e}, Yoann and Fisicaro, Giuseppe and Rolland, Dorian and Genovese, Luigi},
  journal={arXiv:2604.22571},
  year={2026}
}

@Article{MacKnight2025,
author={MacKnight, Robert
and Boiko, Daniil A.
and Regio, Jose Emilio
and Gallegos, Liliana C.
and Neukomm, Th{\'e}o A.
and Gomes, Gabe},
title={Rethinking chemical research in the age of large language models},
journal={Nat. Comput. Sci.},
date={2025-09},
volume={5},
number={9},
pages={715-726},
issn={2662-8457},
doi={10.1038/s43588-025-00811-y},
url={https://doi.org/10.1038/s43588-025-00811-y}
}

@article{Boiko2023Autonomous,
  author  = {Boiko, Daniil A. and MacKnight, Robert and Kline, Ben and Gomes, Gabe},
  title   = {Autonomous chemical research with large language models},
  journal = {Nature},
  volume  = {624},
  number  = {7992},
  pages   = {570--578},
  date    = {2023-12},
  doi     = {10.1038/s41586-023-06792-0}
}

@article{PyzerKnapp2022Accelerating,
  author  = {Pyzer-Knapp, Edward O. and Pitera, Jed W. and Staar, Peter W. J. and Takeda, Seiji and Laino, Teodoro and Sanders, Daniel P. and Sexton, James and Smith, John R. and Curioni, Alessandro},
  title   = {Accelerating materials discovery using artificial intelligence, high performance computing and robotics},
  journal = {npj Computational Materials},
  volume  = {8},
  number  = {1},
  pages   = {84},
  date    = {2022-04},
  doi     = {10.1038/s41524-022-00765-z}
}

@inproceedings{razakh2025multiscale,
author = {Razakh, Taufeq Mohammed and Linker, Thomas and Luo, Ye and Piroozan, Nariman and Pennycook, John and Kumar, Nalini and Musaelian, Albert and Johansson, Anders and Kozinsky, Boris and Kalia, Rajiv K. and Vashishta, Priya and Shimojo, Fuyuki and Hattori, Shinnosuke and Nomura, Ken-ichi and Nakano, Aiichiro},
title = {Multiscale Light-Matter Dynamics in Quantum Materials: From Electrons to Topological Superlattices},
year = {2025},
publisher = {Association for Computing Machinery},
address = {New York, NY, USA},
url = {https://doi.org/10.1145/3712285.3771785},
doi = {10.1145/3712285.3771785},
booktitle = {Proceedings of the International Conference for High Performance Computing, Networking, Storage and Analysis},
pages = {36–47},
numpages = {12},
location = {
},
series = {SC '25}
}

@online{hattori2025beyond,
  author      = {Hattori, Shinnosuke and Shimamura, Kohei and Nakano, Aiichiro and Kalia, Rajiv K. and Vashishta, Priya and Nomura, Ken-ichi},
  title       = {Beyond Scaling: Chemical Intuition as Emergent Ability of Universal Machine Learning Interatomic Potentials},
  date        = {2025-06},
  eprint      = {2506.07579},
  eprinttype  = {arXiv},
  eprintclass = {cond-mat.mtrl-sci},
  url         = {https://arxiv.org/abs/2506.07579}
}

@article{ref:SHIMOJO2019100307,
title = "QXMD: An open-source program for nonadiabatic quantum molecular dynamics",
journal = "SoftwareX",
volume = "10",
pages = "100307",
year = "2019",
issn = "2352-7110",
doi = "10.1016/j.softx.2019.100307",
url = "http://www.sciencedirect.com/science/article/pii/S2352711019300512",
author = "Fuyuki Shimojo and Shogo Fukushima and Hiroyuki Kumazoe and Masaaki Misawa and Satoshi Ohmura and Pankaj Rajak and Kohei Shimamura and Lindsay Bassman and Subodh Tiwari and Rajiv K. Kalia and Aiichiro Nakano and Priya Vashishta"
}

@article{ref:CarPhysRevLett.55.2471,
  title = {Unified Approach for Molecular Dynamics and Density-Functional Theory},
  author = {Car, Roberto and Parrinello, Michele},
  journal = {Physical Review Letters},
  volume = {55},
  issue = {22},
  pages = {2471--2474},
  numpages = {0},
  date = {1985-11},
  publisher = {American Physical Society},
  doi = {10.1103/PhysRevLett.55.2471},
  url = {https://link.aps.org/doi/10.1103/PhysRevLett.55.2471}
}

@article{ref:PayneRevModPhys.64.1045,
  title = {Iterative minimization techniques for ab initio total-energy calculations: molecular dynamics and conjugate gradients},
  author = {Payne, Michael C. and Teter, Michael P. and Allan, Douglas C. and Arias, Tom\'as A. and Joannopoulos, John D.},
  journal = {Reviews of Modern Physics},
  volume = {64},
  issue = {4},
  pages = {1045--1097},
  numpages = {0},
  date = {1992-10},
  publisher = {American Physical Society},
  doi = {10.1103/RevModPhys.64.1045},
  url = {https://link.aps.org/doi/10.1103/RevModPhys.64.1045}
}

@article{ref:HohenbergPhysRev.136.B864,
  title = {Inhomogeneous Electron Gas},
  author = {Hohenberg, Pierre and Kohn, Walter},
  journal = {Physical Review},
  volume = {136},
  issue = {3B},
  pages = {B864--B871},
  numpages = {0},
  date = {1964-11},
  publisher = {American Physical Society},
  doi = {10.1103/PhysRev.136.B864},
  url = {https://link.aps.org/doi/10.1103/PhysRev.136.B864}
}

@article{ref:BlochlPhysRevB.50.17953,
  title = {Projector augmented-wave method},
  author = {Bl\"ochl, Peter E.},
  journal = {Physical Review B},
  volume = {50},
  issue = {24},
  pages = {17953--17979},
  numpages = {0},
  date = {1994-12},
  publisher = {American Physical Society},
  doi = {10.1103/PhysRevB.50.17953},
  url = {https://link.aps.org/doi/10.1103/PhysRevB.50.17953}
}

@article{ref:PerdewPhysRevLett.77.3865,
  title = {Generalized Gradient Approximation Made Simple},
  author = {Perdew, John P. and Burke, Kieron and Ernzerhof, Matthias},
  journal = {Physical Review Letters},
  volume = {77},
  issue = {18},
  pages = {3865--3868},
  numpages = {0},
  date = {1996-10},
  publisher = {American Physical Society},
  doi = {10.1103/PhysRevLett.77.3865},
  url = {https://link.aps.org/doi/10.1103/PhysRevLett.77.3865}
}

@article{ref:GrimmeJCC2512,
author = {Grimme, Stefan},
title = {Accurate description of van der Waals complexes by density functional theory including empirical corrections},
journal = {Journal of Computational Chemistry},
volume = {25},
number = {12},
pages = {1463-1473},
doi = {10.1002/jcc.20078},
url = {https://onlinelibrary.wiley.com/doi/abs/10.1002/jcc.20078},
year = {2004}
}

@article{ref:SunJPC115,
author = {Sun, Tao and Fabris, Stefano and Baroni, Stefano},
title = {Surface Precursors and Reaction Mechanisms for the Thermal Reduction of Graphene Basal Surfaces Oxidized by Atomic Oxygen},
journal = {The Journal of Physical Chemistry C},
volume = {115},
number = {11},
pages = {4730-4737},
year = {2011},
doi = {10.1021/jp111372k},

URL = { 
    
        https://doi.org/10.1021/jp111372k
    
    

},
eprint = { 
    
        https://doi.org/10.1021/jp111372k
    
    

}
}

@article{ref:LarcipreteJACS133,
author = {Larciprete, Rosanna and Fabris, Stefano and Sun, Tao and Lacovig, Paolo and Baraldi, Alessandro and Lizzit, Silvano},
title = {Dual Path Mechanism in the Thermal Reduction of Graphene Oxide},
journal = {Journal of the American Chemical Society},
volume = {133},
number = {43},
pages = {17315-17321},
year = {2011},
doi = {10.1021/ja205168x},
%note ={PMID: 21846143},
URL = {https://doi.org/10.1021/ja205168x},
eprint = {https://doi.org/10.1021/ja205168x}
}

@article{
doi:10.1126/science.aeb9934,
author = {Jian Zhao  and Cameron S. Jorgensen  and Krishnamurthy Mahalingam  and Cynthia Bowers  and Wataru Sugimoto  and Kai Ito  and Seung Ju Kim  and Ruoyu Zhao  and Yichun Xu  and Han-Ting Liao  and Rajiv K. Kalia  and Aiichiro Nakano  and Kohei Shimamura  and Fuyuki Shimojo  and Priya Vashishta  and Ajit K. Roy  and Ning Ge  and Miao Hu  and R. Stanley Williams  and Qiangfei Xia  and Sabyasachi Ganguli  and J. Joshua Yang },
title = {High-temperature memristors enabled by interfacial engineering},
journal = {Science},
volume = {392},
number = {6799},
pages = {771-779},
year = {2026},
doi = {10.1126/science.aeb9934},
URL = {https://www.science.org/doi/abs/10.1126/science.aeb9934},
eprint = {https://www.science.org/doi/pdf/10.1126/science.aeb9934},
}

@article{doi:10.1021/acs.iecr.9b04625,
author = {Kumar, Colonel Vijay and Kandasubramanian, Balasubramanian},
title = {Advances in Ablative Composites of Carbon Based Materials: A Review},
journal = {Industrial \& Engineering Chemistry Research},
volume = {58},
number = {51},
pages = {22663-22701},
year = {2019},
doi = {10.1021/acs.iecr.9b04625},
URL = {https://doi.org/10.1021/acs.iecr.9b04625},
eprint = {https://doi.org/10.1021/acs.iecr.9b04625}
}

@online{nobelprize01,
author ={NobelPrize.org},
year = {2026},
title ={The {N}obel {P}rize in {C}hemistry 2013},
url ={https://www.nobelprize.org/prizes/chemistry/2013/summary},
lastaccessed ={Jun 25, 2026},
}

@INPROCEEDINGS{4090197,
  author={Takemiya, Hiroshi and Tanaka, Yoshio and Sekiguchi, Satoshi and Ogata, Shuji and Kalia, Rajiv K. and Nakano, Aiichiro and Vashishta, Priya},
  booktitle={SC '06: Proceedings of the 2006 ACM/IEEE Conference on Supercomputing}, 
  title={Sustainable Adaptive Grid Supercomputing: Multiscale Simulation of Semiconductor Processing across the Pacific}, 
  year={2006},
  volume={},
  number={},
  pages={23-23},
  doi={10.1109/SC.2006.59}}

@article{Batzner2022NequIP,
  author  = {Batzner, Simon and Musaelian, Albert and Sun, Lixin and Geiger, Mario and Mailoa, Jonathan P. and Kornbluth, Mordechai and Molinari, Nicola and Smidt, Tess E. and Kozinsky, Boris},
  title   = {{E(3)-Equivariant Graph Neural Networks for Data-Efficient and Accurate Interatomic Potentials}},
  journal = {Nature Communications},
  year    = {2022},
  volume  = {13},
  pages   = {2453},
  doi     = {10.1038/s41467-022-29939-5}
}

@inproceedings{Batatia2022MACE,
  author    = {Batatia, Ilyes and Kov\'acs, D\'avid P\'eter and Simm, Gregor N. C. and Ortner, Christoph and Cs\'anyi, G\'abor},
  title     = {{MACE: Higher Order Equivariant Message Passing Neural Networks for Fast and Accurate Force Fields}},
  booktitle = {Advances in Neural Information Processing Systems},
  volume     = {35},
  pages      = {11423--11436},
  year       = {2022}
}

@article{Musaelian2023Allegro,
  author  = {Musaelian, Albert and Batzner, Simon and Johansson, Anders and Sun, Lixin and Owen, Cameron J. and Kornbluth, Mordechai and Kozinsky, Boris},
  title   = {Learning Local Equivariant Representations for Large-Scale Atomistic Dynamics},
  journal = {Nature Communications},
  year    = {2023},
  volume  = {14},
  pages   = {579},
  doi     = {10.1038/s41467-023-36329-y}
}

@article{Batatia2025MACEFoundation,
  author  = {Batatia, Ilyes and Benner, Philipp and Chiang, Yuan and Elena, Adrian M. and Kov\'acs, D\'avid P\'eter and others},
  title   = {A Foundation Model for Atomistic Materials Chemistry},
  journal = {The Journal of Chemical Physics},
  year    = {2025},
  volume  = {163},
  number  = {18},
  pages   = {184110}
}

@article{batatia2025design,
  author    = {Batatia, Ilyes and Batzner, Simon and Kov{\'a}cs, D{\'a}vid P. and Musaelian, Albert and Simm, Gregor N. C. and Drautz, Ralf and Ortner, Christoph and Kozinsky, Boris and Cs{\'a}nyi, G{\'a}bor},
  title     = {The design space of ${E}(3)$-equivariant atom-centred interatomic potentials},
  journal   = {Nature Machine Intelligence},
  year      = {2025},
  volume    = {7},
  number    = {1},
  pages     = {56--67},
  doi       = {10.1038/s42256-024-00956-x},
  url       = {https://www.nature.com/articles/s42256-024-00956-x}
}

@article{riebesell2023matbench,
  author        = {Riebesell, Janosh and Goodall, Rhys E. A. and Benner, Philipp and Chiang, Yuan and Deng, Bowen and Ceder, Gerbrand and Asta, Mark and Lee, Alpha A. and Jain, Anubhav and Persson, Kristin A.},
  title         = {{Matbench Discovery} -- A framework to evaluate machine learning crystal stability predictions},
  journal       = {arXiv preprint arXiv:2308.14920},
  year          = {2023},
  month         = {August},
  eprint        = {2308.14920},
  archivePrefix = {arXiv},
  primaryClass  = {cond-mat.mtrl-sci},
  doi           = {10.48550/arXiv.2308.14920},
  url           = {https://arxiv.org/abs/2308.14920}
}

@article{Drautz2019atomic,
  author    = {Drautz, Ralf},
  title     = {Atomic cluster expansion for accurate and transferable interatomic potentials},
  journal   = {Physical Review B},
  volume    = {99},
  number    = {1},
  pages     = {014104},
  year      = {2019},
  month     = {January},
  publisher = {American Physical Society (APS)},
  doi       = {10.1103/PhysRevB.99.014104},
  url       = {https://link.aps.org/doi/10.1103/PhysRevB.99.014104}
}

@article{Fuetal2022,
  author        = {Fu, Xiang and Wu, Zhenghao and Wang, Wujie and Xie, Tian and Keten, Sinan and Gomez-Bombarelli, Rafael and Jaakkola, Tommi},
  title         = {Forces are not Enough: Benchmark and Critical Evaluation for Machine Learning Force Fields with Molecular Simulations},
  journal       = {Transactions on Machine Learning Research},
  year          = {2023},
  eprint        = {2210.07237},
  archivePrefix = {arXiv},
  primaryClass  = {physics.comp-ph},
  doi           = {10.48550/arXiv.2210.07237},
  url           = {https://arxiv.org/abs/2210.07237}
}

@article{zhang2024dpa2,
  title = {{DPA-2}: A Large Atomic Model as a Multi-Task Learner},
  author = {Zhang, Duo and Liu, Xinzijian and Zhang, Xiangyu and others},
  journal = {npj Computational Materials},
  volume = {10},
  pages = {293},
  year = {2024},
  doi = {10.1038/s41524-024-01493-2},
  archivePrefix = {arXiv},
  eprint = {2312.15492}
}

@misc{zhang2025dpa3,
  title = {A Graph Neural Network for the Era of Large Atomistic Models},
  author = {Zhang, Duo and Peng, Anyang and Cai, Chun and Li, Wentao and Zhou, Yuanchang and Zeng, Jinzhe and Guo, Mingyu and Zhang, Chengqian and Li, Bowen and Jiang, Hong and Zhu, Tong and Jia, Weile and Zhang, Linfeng and Wang, Han},
  year = {2025},
  archivePrefix = {arXiv},
  eprint = {2506.01686},
  doi = {10.48550/arXiv.2506.01686}
}

@misc{liao2026equiformerv3,
  title = {{EquiformerV3}: Scaling Efficient, Expressive, and General {SE(3)}-Equivariant Graph Attention Transformers},
  author = {Liao, Yi-Lun and Hoffman, Alexander J. and Shen, Sabrina C. and Duval, Alexandre and Norwood, Sam Walton and Smidt, Tess},
  year = {2026},
  archivePrefix = {arXiv},
  eprint = {2604.09130},
  doi = {10.48550/arXiv.2604.09130}
}

@misc{mazitov2025petmad,
  title = {{PET-MAD}, a Lightweight Universal Interatomic Potential for Advanced Materials Modeling},
  author = {Mazitov, Arslan and Bigi, Filippo and Kellner, Matthias and Pegolo, Paolo and Tisi, Davide and Fraux, Guillaume and Pozdnyakov, Sergey and Loche, Philip and Ceriotti, Michele},
  year = {2025},
  archivePrefix = {arXiv},
  eprint = {2503.14118},
  doi = {10.48550/arXiv.2503.14118}
}

@misc{zhou2026matris,
  title = {{MatRIS}: Toward Reliable and Efficient Pretrained Machine Learning Interatomic Potentials},
  author = {Zhou, Yuanchang and Hu, Siyu and Zhang, Xiangyu and Wang, Hongyu and Tan, Guangming and Jia, Weile},
  year = {2026},
  archivePrefix = {arXiv},
  eprint = {2603.02002},
  doi = {10.48550/arXiv.2603.02002}
}


\end{document}